# RIBBONS: Rapid Inpainting Based on Browsing of Neighborhood Statisitics

Mojtaba Akbari
Department of Electrical and Computer Engineering
Isfahan University of Technology
Isfahan, Iran
Email: mojtaba.akbari@ec.iut.ac.ir

Majid Mohrekesh
Department of Electrical and Computer Engineering
Isfahan University of Technology
Isfahan, Iran
Email: mmohrekesh@yahoo.com

Nader Karimi
Department of Electrical and Computer Engineering
Isfahan University of Technology
Isfahan, Iran
Email: nader.karimi@cc.iut.ac.ir

Shadrokh Samavi
Department of Electrical and Computer Engineering
Isfahan University of Technology
Isfahan, Iran
Email: samavi96@cc.iut.ac.ir

*Abstract*—Image inpainting refers to filling missing places in images using neighboring pixels. It also has many applications in different tasks of image processing. Most of these applications enhance the image quality by significant unwanted changes or even elimination of some existing pixels. These changes require considerable computational complexities which in turn results in remarkable processing time. In this paper we propose a fast inpainting algorithm called *RIBBONS* based on selection of patches around each missing pixel. This would accelerate the execution speed and the capability of online frame inpainting in video. The applied cost-function is a combination of statistical and spatial features in all neighboring pixels. We evaluate some candidate patches using the proposed cost function and minimize it to achieve the final patch. Experimental results show the higher speed of *RIBBONS* method in comparison with previous methods while being comparable in terms of PSNR and SSIM for the images in MISC dataset.

I. INTRODUCTION

Image inpainting refers to the process of filling missed or corrupted pixels in images using the best approximation based on approximation methods such as the interpolation of neighboring pixels or using the analysis of image texture. It has many applications in various tasks, such as removing undesired objects, or enhancing the image quality especially in old images, and repairing of drop outs on frames of old video magnetic tapes. The work in [1] reviews and compares different inpainting methods and describes them using mathematical equations. There are several kinds of methods proposed for inpainting such as partial differential equation (PDE)-based, exemplar-based, dictionary-based and some other novel methods. In this paper we use image inpainting for removing defects that are in the form of thick missing lines. Such lines could occur in scanned version of old images that are stored for a long period of time.

The work in [2] is one of the pioneers in inpainting which evaluate its work by manipulating an image and re-enhances degraded version to achieve the original document. Its authors used PDE-based image inpainting for removing text from images. The idea was based on propagation of neighbor region intensities into the target which is often the approach of PDE-based image inpainting in anisotropic diffusion of neighbor pixels for reconstruction of missed patch. The major deficiency of [2] was the smoothed or faded edges in the inpainted region that later in 2002, [3] overcomed this challenge with usage of total variation function in the inpainting model to preserve the edges. The PDE-based methods had some drawbacks in natural images because of complex structures and textures of those natural images. Thus, exemplar based algorithms were proposed to make artificial texture based on neighborhood of target region. The authors of [4] proposed exemplar based image inpainting algorithm based on minimization of cost function according to all candidate patches outside of target region. Their proposed algorithm uses iterative scheme and calculates priority and confidence for each candidate patch and finally minimizes the cost function which is a composition of these two parameters to find the best patch. The work in [5] improves proposed algorithm in [4] by changing equation for both priority and confidence by adding a term for curvature. An exhaustive search in all candidates requires a remarkable amount of time and the time will further increases when the process iterates repeatedly. In [6], a multilayer image inpainting method is used which is based on classification of images into layers of similar intensity. Then inpainting algorithm is applied on each layer and the separated layers finally integrate to generate inpainted result. The algorithm of [7] is similar to the method proposed in [5] except the novel method for region filling based on shapes of objects in the neighborhood. It's obvious that [7] is slow as described for [5]. The work in [8] is used for inpainting in order to satisfy the neighborhood coherency of missed patches. The exhaustive search for appropriate match and complexity of matching functions causes a time consuming process to fulfill the inpainting of the image.

Some inpainting methods are based on sparse modeling and minimization of cost functions. The algorithm of [9] uses both total variation, as the edge preserving tool, and $l_0$ normalization of the neighbor pixels in order to estimate missing pixels of the image. So, the estimation could be solved using iterative approaches when it is a minimization problem. The authors of [9] claim that their method is robust against impulse and Gaussian noises but still slow which is result of iteration. The algorithm of [10] uses matrix low rank estimation and sparse representation for patch based image inpainting and shows that their proposed algorithm is robust against impulse noise and can be also used for text removing from image. The main purpose in sparse inpainting is to estimate corrupted pixels based on a dictionary and interpolation of the missed pixels with linear combination of the atoms. The method in [11] is based on using of some dictionary estimation algorithms for inpainting and removing of objects in the image.

Learning methods are sometimes more appropriate to solve inpainting problems, especially in presence of heavy corruption of the image. An example was proposed in [12] which uses deep neural network combined with sparse coding for denoising and image inpainting. Convolutional Neural Network (CNN) is an efficient method of eliminating the noise in image inpainting. The method proposed in [13] uses fully CNN for image inpainting. The proposed network has three convolutional layers to learn features from corrupted image and their equivalent ground truth data. Similarly, [14] presents a CNN for natural image denoising with 4 hidden layers and 24 feature maps. The works based on CNN require learning phase and are appropriate for the set of images similar to the training set.

Our proposed method, called *RIBBONS*: Rapid Inpainting Based on Browsing of Neighborhood Pixels, is a fast algorithm, but most of above methods are time consuming methods and some of them need a training phase that makes the result dependent to the training set and its statistics. *RIBBONS* method achieves to a high speed of processing for a minor loss of quality. Our proposed inpainting algorithm is based on patch selection with the use of neighbor patches in four main directions. The proposed cost function is based on statistics of neighboring pixels. Our algorithm is evaluated using PSNR and SSIM, which are two well-known image quality assessment criteria. We compare our results with the results from similar works.

The structure of the paper is as follows: In section II the proposed method is presented. In section III experimental results are explained and in section IV concluding remarks are offered.

## II. Proposed Method

In this section we first propose a degradation algorithm which simulates what happens on a paper photo after folding the photo for a while and scanning it [15]. Then, the requirements of the proposed method, the cost function and a low pass filter, are described and the main inpainting module is described at the end of this section.

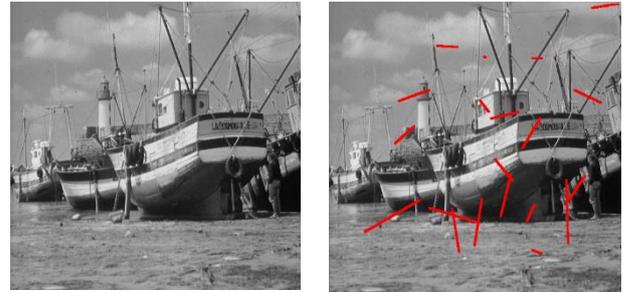

(a)                  (b)

Figure 1: (a) Original image, (b) Degraded image

### A. Image Degradation

Our degradation algorithm produces random ribbons (thick lines) on an image by connecting random couples of pixels. Thus the connecting ribbons have different lengths. A dilation performed on the connecting line ribbons produces desired widths of ribbons. The lengths of ribbons are randomly normal distributed to produce a fair benchmark for testing of the final inpainting algorithm. Different conditions could be achieved as a function of length, width and number of missed ribbons in an image. Figure 1 shows an image and the degraded version with 20 ribbons of 9 pixel width. This random ribbon degradation method can be used to evaluate *RIBBONS* method for different conditions on different images and textures.

### B. Cost Function

*RIBBONS* method is based on finding an appropriate ribbon in the image that is candidate for substitution that minimizes the proposed cost function. The cost function had better represent all features of similarity to select the one that is as best as possible and also acceptable in terms of speed. We employ three features in order to take the similarities between two ribbon patches into account. The features are a combination of simple statistical and spatial information for achieving a powerful means to extract the similarities between ribbons.

The statistical features are the absolute differences between mean and variance between candidate ribbon and neighbor pixels of degraded ribbon which are shown by $\Delta_\mu$ and $\Delta_\sigma$ in Eq. (1). We can choose an equivalent substitute for the ribbon in terms of intensity value and texture by utilizing these statistical measures.

The spatial feature is the distance between center of degraded ribbon and candidate substitute. Our proposed ribbon selection method finds a substitute that is near to degradation region using the spatial metric and also results in more similarity in texture and higher visual transparency of inpainting method output.

Our proposed cost function in Eq. (1) finds an appropriate substitute patch that is similar to region of degradation using three main features that introduced above. We choose the selected ribbon considering this cost function and replace it into target ribbon which was missed or degraded. The best ribbon in our method has the lowest amount of cost among the candidate set.

$$f_{cost} = \Delta_\mu \times \Delta_\sigma \times d \quad (1)$$

Term $d$ represents the distance as mentioned above to consider the spatial feature. Multiplying the costs causes direct effect of changes in any parameter in the final result. For example, multiplying any term by two or reducing 5% of each directly appears in the cost term.

*C. Low-Pass Filtering*

Replacing selected ribbon with the degraded one, causes unwanted edges that need to be eliminated or weakened. Our approach is locally edge smoothing in the boundaries of replaced ribbon which maintains the texture in the inner part of the ribbon. Thus, the boundary edge is eliminated and the inner edges maintains alongside the ribbon.

*D. Proposed Inpainting Scheme*

Now, we put all modules together and describe what is happening in our proposed *RIBBONS* method. The algorithm has four main steps which are shown in Figure 2 and is as follows:

1. Find the places of degradation in main image of RGB color space using the degradation mask that points to the missed ribbons. Each ribbon is now a connected component on the degradation mask.
2. Seek for candidate ribbons without overlapping with the target ribbon in the up, down, right and left direction which may be the best near opportunities. Thus, there are a maximum of four candidate ribbons appropriate for substitution with the missed ribbon. There may be many candidates except these four, but these are chosen for decreasing the complexity of the algorithm while increasing the speed to a rapid inpainting scheme.
3. Select the final candidate, the one with the minimum cost according to the cost function of Eq. (1) and place it in the position of missed ribbon.
4. Apply a Gaussian filter as the low pass filter on the border of missed ribbon and the outer layer after this border which is the border line that is connecting to ribbon part.
5. Select the next ribbon if there is any or end the algorithm if there is no other.

The speed of algorithm is result of substitution rather than making the result pixel by pixel and also selection between limited substitute ribbons. The selection of patches in vertical and horizontal directions results in a good output especially when the frame angle is aligned with that of the buildings and structures in natural life.

## III. EXPERIMENTAL RESULTS

We use MISC database to evaluate *RIBBONS* method. Size of images are $512 \times 512$ and the dataset is available in [16]. The used platform for comparison of our method with other preproposed methods is a laptop system which has a Corei7 CPU with 6GB of RAM memory. The codes are implemented on MATLAB R2014a translator without using any MEX file for our codes although the most similar works to ours, [5] and [8], are using it to accelerate the speed of execution. This fact demonstrates our further code rapidness in *RIBBONS* method. Table I. shows that our method is much faster than the methods proposed in [5] and [8], whereas both methods have comparable level of degradation. The execution could even be faster with using code vectorization and GPU usage for increasing the potential of online processing in applications such as video enhancement.

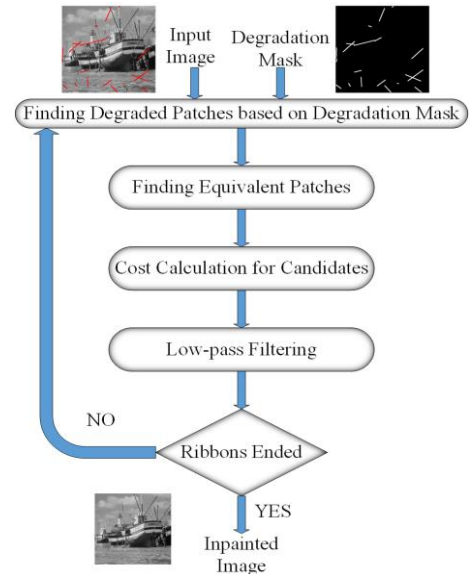

Figure 2: Block diagram of *RIBBONS* method

Table I. Comparison of execution time on "*Boat*", "*Lena*"& "*Goldhill*" images

| Image | Criterion | Criminisi [5] | Korman [8] | RIBBONS |
|---|---|---|---|---|
| Boat | PSNR | 38.23 | 39.31 | 36.32 |
|  | SSIM | 0.99 | 0.99 | 0.99 |
|  | TIME (sec) | 71.60 | 93.18 | **1.78** |
| Lena | PSNR | 41.45 | 41.55 | 38.3 |
|  | SSIM | 0.99 | 0.99 | 0.99 |
|  | TIME (sec) | 63.28 | 74.61 | **1.20** |
| Goldhill | PSNR | 38.24 | 39.62 | 38.51 |
|  | SSIM | 0.98 | 0.98 | 0.98 |
|  | TIME (sec) | 69.85 | 41.95 | **1.68** |

Figure 3 shows *RIBBONS* method results for different line widths which demonstrates an average approximate enhancement of 52.6% improvement in PSNR and 3.22% in SSIM. We can also see that our inpainting method removes most of artifacts in degraded image.

Figure 4 shows variation of PSNR for all images in MISC dataset. Each box in the chart shows the set of PSNR values resulted after inpainting for a specific degradation in all images.

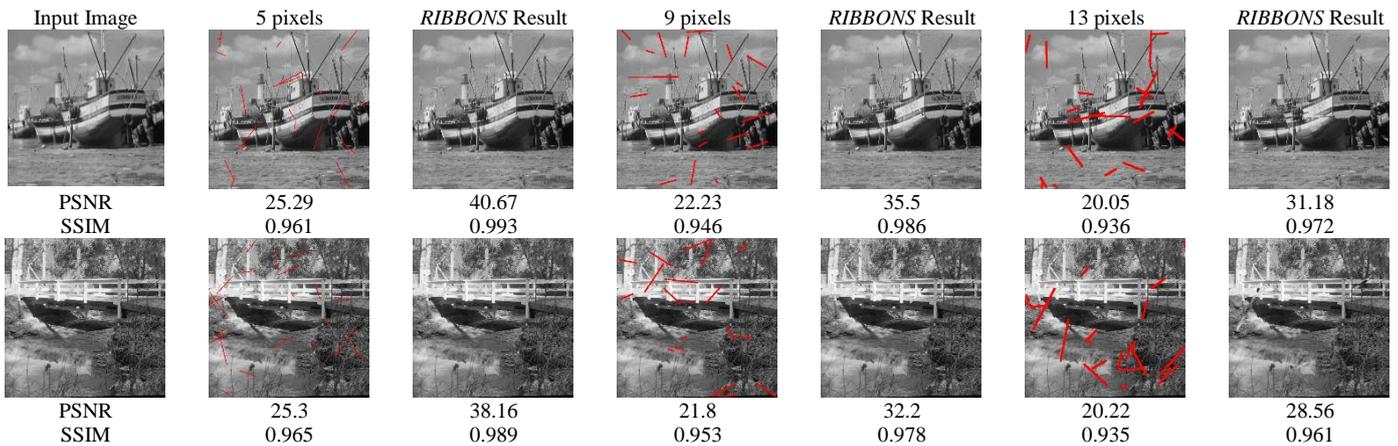

Figure 3: *RIBBONS* method results for different line widths of degradation

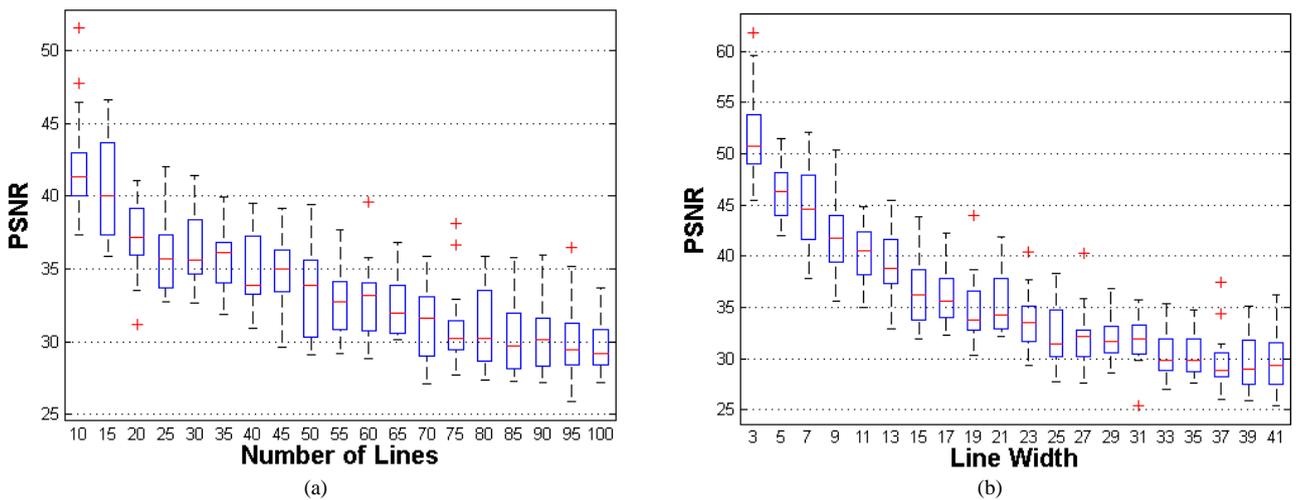

Figure 4: Deviation of PSNR for all images of MISC dataset, (a) Different number of lines, (b) Different line widths

Figure 4(a) shows the box plot for different number of nine-pixel-width scratching lines and figure 4(b) demonstrates that of 10 lines of different widths. The selection of ribbon adjacent to target, results in rather appropriate quality of the image as it is approved by the charts of figure 4 for various image textures and objects. The limited number of outlier values shows the uniformity of algorithm in different situations although most of such instances are upper than the main body and demonstrate extra better results in some conditions. Figure 4 shows that decreasing the number of degraded pixels results in superior quality of enhancement.

We also evaluate SSIM results of *RIBBONS* method for different widths and number of lines while degrading the image quality. Figure 5 shows SSIM result of inpainted image for different degradation parameters in the images of Lena and Barbara. Again, as in Figure 4, decreasing the number of lines or line widths causes overall increasing of SSIM in Figure 5. The deviation from monotonically reduction in both Figures 4 and 5, is result of random line selection in our code simulation for degradation.

IV. CONCLUSION

In this paper we proposed a novel rapid image inpainting algorithm, called *RIBBONS* method based on patch selection around each connected component of degradation. *RIBBONS* method uses a cost function to evaluate each candidate patch and selects one that minimizes the cost function. Then we use a low-pass filter to smooth edges that will occur in the first stage of inpainting. *RIBBONS* method is also fast in comparison to other similar inpainting methods and has a good result in different images with different rate of degradation. We also evaluate proposed method for different images in MISC dataset.

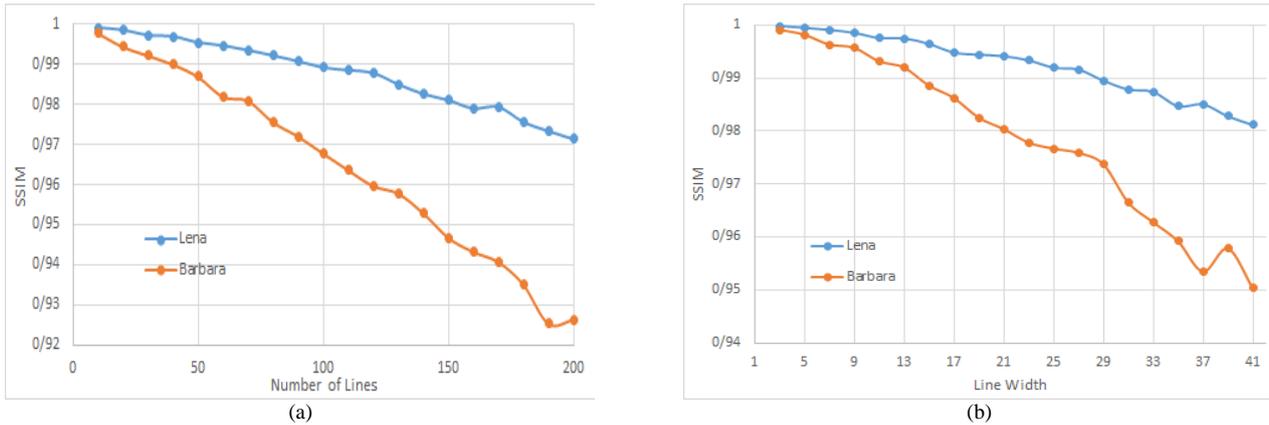

Figure 5: Deviation of SSIM for images of Lena and Barbara, (a) Different number of lines, (b) Different line widths